\documentclass[twocolumn,amssymb,amsmath,showpacs]{revtex4}

\usepackage{graphicx}
\usepackage{bbm}

\newcommand{\mrm}{\mathrm}
\newcommand{\mbf}{\mathbf}

\newcommand{\im}{\mathrm{i}}
\newcommand{\de}{\mathrm{d}}
\newcommand{\GL}{\Gamma_\mathrm{L}}
\newcommand{\GR}{\Gamma_\mathrm{R}}
\newcommand{\GRud}{\Gamma_\mathrm{U/D}}

\newcommand{\kBT}{k_\mathrm{B}T}
\newcommand{\elec}{e}
\newcommand{\hplain}{h}

\begin{document}

\title{Generation and detection of a spin entanglement in nonequilibrium quantum dots}

\author{Stefan Legel$^1$, J\"urgen K\"onig$^2$, and Gerd Sch\"on$^1$}
\address{$^1$Institut f\"ur Theoretische Festk\"orperphysik and DFG-Center for Functional Nanostructures (CFN), Universit\"at Karlsruhe, 76128 Karlsruhe, Germany}
\address{$^2$Institut f\"ur Theoretische Physik III, Ruhr-Universit\"at Bochum, 44780 Bochum, Germany}

\begin{abstract}
Spin entanglement between two spatially separated electrons can be generated in nonequilibrium interacting quantum dots, coherently coupled to a common lead. In this system entangled two-electron states  develop which are Werner states with an imbalance between singlet and triplet probabilities.
We propose a multi-terminal, multiply-connected setup for the generation and detection of this imbalance.
In particular, we identify a regime in which the formation of spin entanglement
leads to a cancellation of Aharonov-Bohm oscillations. 
\end{abstract}


\maketitle

\section{Introduction}

The controlled generation and detection of entanglement of quantum states 
remains one of the fundamental challenges of quantum physics.
Experiments with entangled photons have already entered the realm of advanced quantum communication and cryptography. 
In solid-state systems electron spins are considered as prime candidates for 
the demonstration of such effects. 
Spin degrees of freedom are only weakly coupled to the environment, which leads to long decoherence times \cite{KoppensScience309,PettaScience309}, and coherence lengths well exceeding the micrometer range.
Several setups involving quantum dots have been proposed to create and detect pairs of spatially separated, spin-entangled electrons.
The schemes rely, e.g., on the extraction of Cooper pairs from a superconductor \cite{RecherSukhorukovPRB63}, or the separation of a pair of entangled electrons from a singlet ground state of single \cite{OliverPRL88,SaragaLossPRL90} or double quantum dots \cite{HuDasSarmaPRB69,BlaauboerPRL95}.   
Evidence of entanglement in these systems can be obtained from noise measurements \cite{BurkardLossPRB61} or coincidence detection \cite{BlaauboerPRL95}.
In Ref.~\cite{LossSukhorukovPRL84} a detection scheme was proposed where pure singlet and triplet states can be distinguished by Aharonov-Bohm interferometry.

\begin{figure}
 \begin{center}
   \includegraphics[scale=0.5]{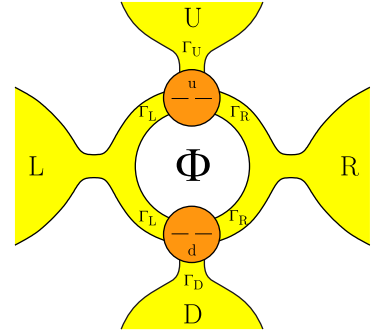}
 \end{center}
  \caption{ \label{Fig:ABI-Fork} Two quantum dots (u and d) coupled to a joint reservoir on the left hand side (L) and to two separate leads at the top and the bottom  (U and D). The second joint reservoir on the right (R) closes the Aharonov-Bohm ring and is used to probe the state of the system. }
\end{figure}

In this article we analyze a setup with two spatially separated quantum dots,
illustrated in Fig.~\ref{Fig:ABI-Fork}, which allows the creation and detection of entanglement of spatially separated electron spins. The two dots are coupled coherently to a joint source electrode on the left and to two independent drain reservoirs at the top and bottom.
When a bias voltage is applied, electrons are driven from the source via the dots to the drain electrodes. 
A nonequilibrium, mixed electronic state is created. 
For appropriate values of the voltage, due to the strong onsite Coulomb repulsion, a state with two electrons (one on each dot) has a high probability.
It turns out to be of the form of a Werner state, with a
strong enhancement of the singlet 
component,
although singlet and triplet states are degenerate \cite{LegelKoenig}.

In order to detect the 
entanglement, an
 additional joint reservoir, closing the Aharonov-Bohm geometry, is coupled to the system. The current to this reservoir is studied in  the cotunneling regime.
Under certain conditions the entanglement leads to a suppression of the Aharonov-Bohm oscillations. Hence this part of the setup serves as probe of the state of the system and as detector for entanglement.
Exploring the results in a wide parameter 
range
we identify regimes where the double dot has a large probability to be in a singlet state. 
The predicted behavior provides a proof of concept for the entanglement generation in coherently coupled, nonequilibrium quantum dots.

\section{Model}

We consider the system shown in Figure \ref{Fig:ABI-Fork}. 
It consists of two single-level quantum dots, $i=\mrm{u,d}$, with strong on-site Coulomb repulsion $U$ described by the Hamiltonian 
\begin{equation}
H_\mrm{dots} = \sum_{i\sigma}  \varepsilon_{i} \, c_{i\sigma}^\dagger c_{i\sigma} + \sum_{i} U \, c_{i\uparrow}^\dagger c_{i\downarrow}^\dagger c_{i\downarrow} c_{i\uparrow}  \, .
\end{equation}
The dots are coupled to joint reservoirs on the left and right forming an Aharonov-Bohm loop, and to two separate leads ``up'' and ``down''. The total system is then modelled by 
$H =  H_\mrm{dots} + H_\mrm{res} + H_\mrm{t}$ with the four leads, $r =\mrm{L,R,U,D}$, with spin degenerate energies $\varepsilon_{rk}$,
\begin{equation}
H_\mrm{res}=  \sum_{rk\sigma} \varepsilon_{rk} \, a_{rk\sigma}^\dagger  a_{rk\sigma} \, ,
\end{equation} 
and the four tunnel Hamiltonians  $H_\mrm{t}= H_\mrm{tL} + H_\mrm{tR} + H_\mrm{tU} + H_\mrm{tD}$. 
Tunneling between the upper dot and reservoir is simply described by
\begin{equation}
  H_{\mrm{tU}} 
    = \sum_{k\sigma} \left[ t_{U} \, c^\dagger_{u\sigma} a_{Uk\sigma} + \mrm{h.c.} \right] ,
\end{equation}
and similar for the lower pair.
Tunneling between the joint leads on the left and the right, on the other hand, is sensitive to the phases of the matrix elements. Without loss of generality we can distribute them in a symmetric gauge, 
\begin{equation}
\begin{split}
  H_{\mrm{tL}} 
    & = \sum_{k\sigma} \left[  \left( t_\mrm{Lu} \mrm{e}^{-\im \varphi/4} \, c^\dagger_{\mrm{u}\sigma} + t_\mrm{Ld} \mrm{e}^{+\im \varphi/4} \, c^\dagger_{\mrm{d}\sigma} \right) a_{\mrm{L}k\sigma} + \mrm{h.c.} \right] ,
    \\ 
  H_{\mrm{tR}} 
    & = \sum_{k\sigma} \left[ \left( t_\mrm{Ru} \mrm{e}^{+\im \varphi/4} \, c^\dagger_{\mrm{u}\sigma} + t_\mrm{Rd} \mrm{e}^{-\im \varphi/4} \, c^\dagger_{\mrm{d}\sigma} \right) a_{\mrm{R}k\sigma} + \mrm{h.c.} \right] .
\label{Eq:ABI_Ht}
\end{split}
\end{equation}
All tunneling matrix elements are taken to be independent of spin and energy, and they can be chosen real and positive.
The leads are assumed to be in local equilibrium, described by electrochemical potentials $\mu_r$. 

To keep the discussion transparent, we assume some symmetries. We set the upper and the lower reservoir to the same potential, $\mu_\mrm{U/D} = \mu_\mrm{U} = \mu_\mrm{D}$, and 
restrict ourselves to degenerate dot levels, $\varepsilon = \varepsilon_\mrm{u} = \varepsilon_\mrm{d}$.
We, furthermore, assume symmetric coupling strengths of the upper and lower dot to the left reservoir, $t_\mrm{Lu}= t_\mrm{Ld}= t_\mrm{L}$, and similar for the right reservoir, $t_\mrm{Ru}= t_\mrm{Rd}= t_\mrm{R}$, and we assume 
$t_\mrm{U}= t_\mrm{D}= t_\mrm{U/D}$.
Defining the tunneling strength by $\Gamma_{r} = 2\pi \, t_{r}^2 \, N_r $, where $N_r$ is the density of states of the relevant reservoir, we are left with three parameters $\GL$, $\GR$, and $\GRud$, 
which gives us enough freedom to tune the system into the interesting regimes. 
Double occupancy of each dot is suppressed by a strong Coulomb repulsion $U$, for simplicity assumed here to be larger than all other energy scales. 
However, the generalization to finite $U$ is straightforward.
We apply two bias voltages measured relative to $\mu_\mrm{L} = 0$: 
a strong bias voltage between the left and the upper and lower reservoirs, 
$\elec V_\mrm{U/D}=\mu_\mrm{U/D}$,
and a small transport voltage between the left and right reservoir, $\elec V_\mrm{R}=\mu_\mrm{R}$, used to probe the state of the system.

In essence, the preparation and measurement scheme analyzed in the following is based on two observations. First, the components of the density matrix which describe the doubly occupied system with one electron on each dot is a so-called Werner state, a mixture of a singlet and three equivalent triplet states. The distribution between singlet and triplet probabilities can be tuned in a certain range by the coupling strengths and the potentials applied to the system. For illustration we consider the state obtained immediately after charging the double dot with two electrons with opposite spin from the left reservoir
\begin{eqnarray}
( e^{-\im\varphi/4} \, c^\dagger_{\mrm{u}\sigma} + 
e^{+\im\varphi/4} \, c^\dagger_{\mrm{d}\sigma} )
( e^{-\im\varphi/4} \, c^\dagger_{\mrm{u}\bar{\sigma}} 
+ e^{+\im\varphi/4} \, c^\dagger_{\mrm{d}\bar{\sigma}} ) |0\rangle
\nonumber\\
=
e^{-\im\varphi/2} |\sigma\bar{\sigma},0\rangle + 
e^{+\im\varphi/2} |0,\sigma\bar{\sigma}\rangle 
+ |\sigma,\bar{\sigma}\rangle - |\bar{\sigma},\sigma\rangle \nonumber
\nonumber \, .
\end{eqnarray}
The minus sign in the last term is a direct consequence of the Fermi statistics. In view of this sign and the 
Pauli
principle it is obvious that the equivalent expression for a pair of electrons with equal spin vanishes. 
A strong onsite Coulomb interaction prohibits double occupancy of each dot. Hence the state reduces to a perfectly entangled, pure singlet state
$$
|\mrm{S}\rangle  = \left(|\sigma,\bar{\sigma}\rangle 
- |\bar{\sigma},\sigma\rangle \right)/\sqrt{2}\, .
$$  
In equilibrium, however, the state has relaxed to a completely unentangled mixed state with uniformly distributed probabilities for singlet and triplets.

The second observation is that for the cotunneling rates the phases of the  transport contributions for singlet and triplet states are shifted relative to each other by $\pi$. Hence they yield different interference patterns in the Aharonov-Bohm interferometer. 

The strategy, therefore, is to use the subsystem composed of the left, upper, and lower reservoirs, and the dots to prepare the state of the system. In particular, we tune the ratio of singlet and triplet probabilities via the bias voltage $V_\mrm{U/D}$. 
The signatures of this state in the transport are 
monitored by the Aharonov-Bohm interferometer, 
which is operated in the cotunneling regime for the 
tunneling to the right reservoir.
The main message of this article is that for a setup with cascading coupling strengths, $\GL \gg \GRud \gg \GR$, an enhanced probability for singlet 
over triplet formation is associated with a suppression of the oscillations in the conductance of the Aharonov-Bohm probe.

\section{Preparation of Stationary Entangled States}\label{Sec:ABI-Fork_PrepStatState}

We investigate the system in the frame of the real-time diagrammatic technique Refs.~\cite{Real-timeDiagrammatic_2,double-dots_2,Pohjola} which allows us 
to go beyond an orthodox master equation approach. In particular we account for the quantum coherent time evolution of the double dot and focus on correlations between the electrons.
The elements of the reduced density matrix are given by 
$p_\chi^{\chi '} = \langle \chi'|\mrm{tr_{res}}\, \rho |\chi\rangle$, 
where the reservoir degrees of freedom of the full density matrix are traced out, and $\chi$ and $\chi'$ label general many-body states of the double dot. 
For strong Coulomb repulsion, the reduced Hilbert space is spanned by nine basis states $|\chi_\mrm{u},\chi_\mrm{d}\rangle$ with $\chi_i =0,\uparrow,\downarrow$.

To describe the symmetries implied by tunneling we introduce a tailored basis. 
The state of the empty system and its probability are denoted by 
$|0,0\rangle$ and $p_0$, 
respectively. 
A single electron of spin $\sigma$ can occupy, with overall probability for single occupancy $p_1 = \sum_{i\sigma} p_{i\sigma}$, either the upper or the lower dot, or, in general, any coherent superposition of these two. Thus, the corresponding block of the density matrix describes a quantum two-state system, which we express an isospin for each spin component, $\mbf{I}_\sigma$.
A natural basis are the two states with one electron in the upper or in the lower dot.
However, in this basis the joint left and right reservoirs couple via the Hamiltonian (\ref{Eq:ABI_Ht}) to the non-diagonal 
components 
$\mbf{I}_\sigma \cdot \mbf{m}_\mrm{L/R}$
of the isospin, with 
\begin{equation}
\mbf{m}_\mrm{L/R} = (\cos\frac{\varphi}{2},\pm\sin\frac{\varphi}{2},0) \, ,
\end{equation}
which introduces a non-diagonal time evolution of the reduced density matrix. Electrons entering the double dot from the left reservoir generate an isospin polarization in direction of $\mbf{m}_\mrm{L}$, whereas electrons entering from the right are polarized in the direction of $\mbf{m}_\mrm{R}$. 
We further note, that due to the spin symmetry of the Hamiltonian the two spin components are completely equivalent. We will formulate equations for the 
expectation values of the isospin for either spin component, 
$\langle \mbf{I}_\uparrow \rangle = \langle \mbf{I}_\downarrow \rangle
= \mbf{I}_{\uparrow / \downarrow}$. To simplify expresssions we dropped in the last form the angular brackets.

The doubly occupied subspace with one electron on each dot is naturally spanned by the spin singlet $|\mrm{S}\rangle$, which we find with probability $p_\mrm{S}$,
and the three triplet states $|\mrm{T}_+\rangle$, 
$|\mrm{T}_0\rangle$,
and 
$|\mrm{T}_-\rangle$.
Due to the spin symmetry of the Hamiltonian the three triplet states are equivalent and are occupied with equal probability $p_\mrm{T_0} = p_\mrm{T_\pm}= p_\mrm{T}/3$. 
The corresponding block of the density matrix can be represented in the Werner form \cite{Werner}
\begin{equation}
  W(F)=F\, |{\rm S}\rangle\langle {\rm S}| + (1-F)\, \frac{\openone_4 -|{\rm S}
    \rangle\langle {\rm S}|}{3} \, ,
\end{equation}
where the coefficient $F = p_\mrm{S}/(p_\mrm{S}+p_\mrm{T})$ denotes the Werner fidelity. Values in the range of $1/2 < F \le 1$ indicate a quantum correlated mixed state from which arbitrary entanglement can be distilled by suitable purification protocols \cite{BennettPRA,Linden}. 
As we will show below, the Werner fidelity of the stationary state can be tuned in a certain range by the tunnel couplings and the applied bias voltage. Regimes with $F > 1/2$ are feasible if the double dot is predominantly charged from a common reservoir.

The kinetic equations for the reduced density matrix can be arranged in vector form with the diagonal probabilities collected in 
$\mbf{p} = (p_0,p_1,p_\mrm{S},p_\mrm{T})$ 
and the state of the singly occupied subspace specified by the isospin 
$\mbf{I}_{\uparrow / \downarrow} = (I_x,I_y,I_z)$.
In lowest order in the coupling strengths $\GL,\GR,\GRud$ we obtain 
\begin{widetext}
\begin{equation}
\begin{split}
  \frac{\de }{\de t} \mbf{p} = & \sum_{r=\mrm{L,R,U/D}} \Gamma_r 
    \begin{pmatrix}
      -4\,f_r  & 1-f_r & 0 & 0 \\
      4\,f_r & -1-f_r & 2-2\,f_r & 2-2\,f_r \\
      0 & f_r/2 & -2 + 2\,f_r & 0 \\
      0 & 3\,f_r/2 & 0 & -2+2\,f_r
    \end{pmatrix}
    \mbf{p} +
    \sum_{r=\mrm{L,R}} \Gamma_r 
    \begin{pmatrix}
      4-4\,f_r \\
      -4+8\,f_r \\
      2\,f_r \\ 
      -6\,f_r
    \end{pmatrix}
  (\mbf{I}_{\uparrow / \downarrow} \cdot \mbf{m}_r) 
\\ \\
  \frac{\de }{\de t} \mbf{I}_{\uparrow / \downarrow} = & \sum_{r=\mrm{L,R}} \Gamma_r \left[
    f_r \,p_0 + \left( f_r - \frac{1}{2} \right) \frac{p_1}{2} + 
    (1-f_r) \,\frac{p_\mrm{S}}{2} - (1-f_r) \,\frac{p_\mrm{T}}{2} \right] \mbf{m}_r -
  \sum_{r=\mrm{L,R,U/D}} \Gamma_r \,(1+f_r) \,\mbf{I}_{\uparrow / \downarrow} \, ,
  \label{Eq:KinStationaryABI-Fork}
\end{split}
\end{equation}
\end{widetext}
where the Fermi distributions $f_r = 1/(e^{\beta(\varepsilon-\mu_r)}+1)$ 
in the respective reservoirs are evaluated at the dots' energy $\varepsilon$. The Equations~(\ref{Eq:KinStationaryABI-Fork}) extend our earlier 
work~\cite{LegelKoenig}, where we considered a double-dot system coupled to three reservoirs, but without the Aharonov-Bohm probe and right reservoir.

To generate and detect a singlet-triplet imbalance, the system parameters have to be chosen specifically.
First, the tunnel coupling to the right lead should be much weaker than the other ones, $\GR \ll \GL, \GRud$, in order to ensure that the state of the double dot is not affected by the measurement via the interferometer. 
Furthermore, only a small bias voltage is applied across the Aharonov-Bohm ring, and the dot levels are kept off resonance such that the flux-sensitive linear conductance is dominated by cotunneling processes.
Finally, in order to find with large probability two electrons in the system, the rate for charging should be much larger than for discharging, $\GL \gg \GRud$.

In the following, we compare two situations:\\
(i) For vanishing bias voltage between the left and upper/lower reservoirs, the double dot remains in equilibrium, and $F = 1/4$.
This defines the reference for comparison.\\
(ii) By applying a high bias voltage between the left and upper/lower reservoirs, we generate a singlet-triplet imbalance with $F > 1/2$.

\section{Probing the Singlet-Triplet Imbalance}

\begin{figure} 
 \begin{center}
   \includegraphics[scale=0.5,clip]{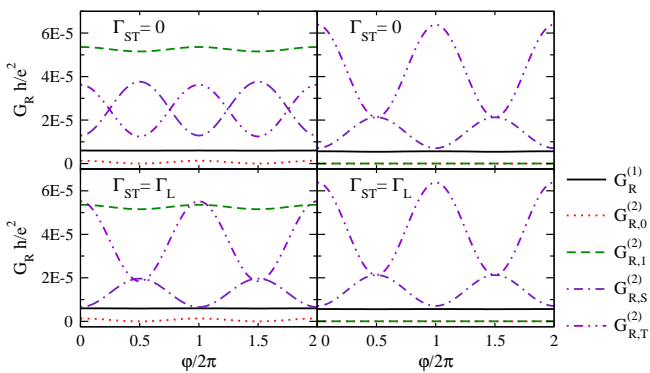}
 \end{center}
 \caption{ \label{Fig:ABI-Fork_GRcomp_phi} The individual contributions to the linear conductance are plotted with and without singlet-triplet relaxation. The first order, $G_\mrm{R}^{(1)}$, and second order terms, $G_{\mrm{R},\chi}^{(2)}$, are specified for equilibrium (right plots) and for strong bias voltage, $\elec V_\mrm{U/D} = -20\,\kBT$ (left plots). The parameters are $\GL = 1\,\kBT$, $\GRud = 0.1\,\kBT$, $\GR = 0.01\,\kBT$, $\varepsilon = -10\,\kBT$ and $|\elec V_\mrm{R}| \ll \kBT$.} 
\end{figure}

Cases (i) and (ii) lead to significantly different
interference signatures in the transport through the Aharonov-Bohm ring.
While the first one shows a strong flux dependence and large amplitude of the Aharonov-Bohm oscillations, they are suppressed in the second case, where the Werner fidelity approaches $1/2$ if the double dot is predominantly occupied with two electrons. 
To 
demonstrate
this difference we calculate the linear conductance through the Aharonov-Bohm interferometer.
The contribution to first order in the coupling strength is evaluated within the diagrammatic technique. It turns out to be small due to our choice of parameters (compare also Figure \ref{Fig:ABI-Fork_GRcomp_phi}).
For the dominant second-order contribution we calculate the stationary density matrix according to (\ref{Eq:KinStationaryABI-Fork}) and obtain the cotunneling rates from Fermi's Golden rule 
\cite{AverinNazarovPRL65,AverinNazarov,Akera}
\begin{equation}
  (2\pi)^2\,N_\mrm{L}N_\mrm{R}\,
  \sum_{\chi_\mrm{f}, k_\mrm{f}} 
  \left| \sum_n \frac{\langle \chi_\mrm{f}\,k_\mrm{f} | H_\mrm{t} | n \rangle\langle n | H_\mrm{t} | \chi_\mrm{i}\,k_\mrm{i} \rangle}{\omega - E_n}
  \right|^2 
\, .
\label{Eq:S_chi_i}
\end{equation}
Here the initial double-dot state is denoted by $\chi_\mrm{i}$, and all possible final states are summed over. 
The energy $E_n$ of the intermediate virtual state $|n\rangle$ is either 0, $\varepsilon$ or $2\varepsilon$, and $\omega$ is the energy of the incoming lead electron.

Gathering all cotunneling terms corresponding to the four scenarios where the dot system is  empty ($\chi=0$), occupied with a single electron with isospin $\mbf{I}_{\uparrow / \downarrow}$ ($\chi=1$), or occupied with two electrons either in a singlet ($\chi=\rm S$) or a triplet ($\chi=\rm T$) state, we find the second order contribution to the linear conductance 
$G_\mrm{R} = \left. \partial I_\mrm{LR}/\partial V_\mrm{R} 
\right|_{V_\mrm{R}=0}$,
\begin{equation}
  G_\mrm{R}^{(2)} = 
  -\frac{\elec^2}{\hplain}\, g (\varphi) \,\mrm{Re} \int\de\omega\, 
  \frac{\GL\GR}{(\omega - \varepsilon + i0^+)^2} \,f'(\omega) \, .
\label{Eq:Glin2Classical}
\end{equation}
The Aharonov-Bohm flux dependence is included in the dimensionless
conductance $g(\varphi) = \sum_\chi g_\chi(\varphi)$, with
\begin{subequations}
\begin{align}
  g_0 (\varphi) &= 2 \,(1+\cos\varphi)\, p_0(\varphi) 
\\ 
  g_1 (\varphi) &= 2 \,(3-\cos\varphi)\, p_1(\varphi) 
    + 16 \,(\mbf{m}_\mrm{L}-\mbf{m}_\mrm{R}) \cdot \mbf{I}_{\uparrow / \downarrow}
\\ 
  g_\mrm{S} (\varphi) &= \left( \frac{1}{2} - \frac{1}{4}\cos\varphi \right)
\, p_\mrm{S}(\varphi)
\label{Eq:g_S}
\\ 
  g_\mrm{T} (\varphi) &= \left( \frac{1}{2} + \frac{1}{4}\cos\varphi \right)
\, p_\mrm{T}(\varphi)
\label{Eq:g_T}
\end{align}
\label{Eq:gs}
\end{subequations}
The flux sensitivity of the conductance is twofold.
First, the phase factors from the tunneling Hamiltonian are taken into account in the coherent summation of processes
in the cotunneling rates (\ref{Eq:S_chi_i}).
In particular, as discussed in \cite{LossSukhorukovPRL84}, the phase dependence of the contributions from an initial singlet and triplet state (Equations (\ref{Eq:g_S}) and (\ref{Eq:g_T})) are  shifted relative to each other by $\pi$.
We emphasize that these results rely on the degeneracy of singlet and triplets.
All four constitute equivalently possible final states of cotunneling processes with an initially doubly occupied system.
If there was a large singlet-triplet energy splitting with transitions between singlet and triplet states being suppressed, then
$g_\mrm{S,T} (\varphi) = (1 + \cos\varphi)\, p_\mrm{S,T}/4$
without any distinction between the contributions.

The second kind of flux dependence results from the stationary state 
distribution of the system. In general nonequilibrium situations, the reduced density matrix, represented by $p_0$, $p_1$, $\mbf{I}_{\uparrow / \downarrow}$, $p_\mrm{S}$, and $p_\mrm{T}$, is influenced by the Aharonov-Bohm probe and becomes phase sensitive.
This effect is weak for weak tunnel coupling $\GR \ll \GL, \GRud$ where the state of the system is only weakly influenced by the AB-probe.
In the limit $\GL \gg \GRud \gg \GR$ and the double dot charged from the left reservoir the isospin points approximately in direction of $\mbf{m}_\mrm{L}$ such that 
$g_1 (\varphi) \approx 2 \,(3-\cos\varphi)\, p_1 - 16 \,(1-\cos\varphi)\, |\mbf{I}_{\uparrow / \downarrow}|$.

We compare now the two scenarios (i) and (ii). For low dot energies in an equilibrium situation, $\varepsilon \ll \mu_r$, the double dot is mainly occupied with two electrons with 
$p_\mrm{T} \approx 3/4$ and $p_\mrm{S}\approx 1/4$. As a consequence, 
in this equilibrium reference situation there is a good visibility of the Aharonov-Bohm oscillations,
\begin{equation} 
  g^{(\mrm{i})}(\varphi) 
  \approx \frac{1}{2} + \frac{1}{8}\cos\varphi \, .
\label{Eq:g_eq}
\end{equation}
In contrast, for strong bias voltage, the phase dependence is suppressed. To see this, we expand the density matrix for a strong asymmetry in the coupling to the left, and the upper and lower reservoirs, $x=\GRud/\GL \ll 1$, and obtain 
$p_0 = {\cal O}(x^2)$, $p_1 = 4x/3 + {\cal O}(x^2)$,
$\mbf{I}_{\uparrow / \downarrow} \approx x/6\,\mbf{m}_\mrm{L} + {\cal O}(x^2)$, 
$p_\mrm{S} = 1/2 - x/2 + {\cal O}(x^2)$, and
$p_\mrm{T} = 1/2 - 5x/6 + {\cal O}(x^2)$.
Thus, the dimensionless conductance approaches
\begin{equation} 
  g^{(\mrm{ii})}(\varphi) 
  \approx \frac{1}{2} + \frac{\GRud}{\GL} 
  \left( \frac{14}{3} - \frac{29}{12} \cos\varphi \right) \, .
\label{Eq:g_noneq}
\end{equation}
For strong asymmetry, $\GRud \ll \GL$, the oscillations of the singlet and triplet terms cancel each other and the phase dependence of the conductance vanishes.
This suppression of the Aharonov-Bohm amplitude  allows us to detect the singlet-triplet asymmetry generated in system.

We remark that if the system is prepared in a state with maximal
entanglement, $p_{\mrm{S}} = 1$ or $p_{\mrm{T}} = 1$, as considered in 
Ref.~\cite{LossSukhorukovPRL84}, then the visibility of 
the Aharonov-Bohm oscillations would be even twice as large as in equilibrium,
$g=\frac{1}{2} \mp \frac{1}{4} \cos \varphi$.
For a steady-state scenario considered in this paper, however, it is not 
possible to have both a large overall probability of double occupancy of the 
double dot and a maximal singlet-triplet imbalance at the same time.
To optimize the probability for double occupancy we choose $\GL \gg \GRud$,
for which $p_{\mrm{S}} = p_{\mrm{T}} = 1/2$, which indicates finite but not
maximal entanglement.

\begin{figure} 
 \begin{center}
   \includegraphics[scale=0.48,clip]{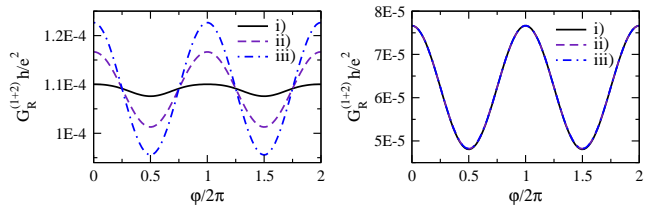}
 \end{center}
 \caption{ \label{Fig:ABI-Fork_GR_phi} The linear conductance of the Aharonov-Bohm subsystem up to second order in the coupling strength is plotted versus the Aharonov-Bohm phase. The parameters are $\GL = 1\,\kBT$, $\GRud = 0.1\,\kBT$, $\GR = 0.01\,\kBT$, $\varepsilon = -10\,\kBT$ and $|\elec V_\mrm{R}| \ll \kBT$. 
For strong bias voltage between the left and the upper/lower reservoirs, $\elec V_\mrm{U/D} = -20\,\kBT$ (left plot), and in equilibrium (right plot), the ideal system (i) is compared to results for singlet-triplet decay rates of (ii) $\Gamma_\mrm{ST} = 0.5\,\GRud$ and (iii) $\Gamma_\mrm{ST} = 10\,\GRud$.} 
\end{figure}

The plots in Figure \ref{Fig:ABI-Fork_GR_phi} display the linear conductance up to second order in the coupling strength based on a full solution of the stationary equations (\ref{Eq:KinStationaryABI-Fork}). 
If the system is close to equilibrium, $|\elec V_\mrm{U/D}| \ll \kBT$, the triplet probability is large ($F \approx 1/4$), and the conductance is dominated by a positive $\cos\varphi$ kind of oscillation (compare Equation (\ref{Eq:g_eq})). 
In contrast, the interference is suppressed if the system is driven into a singlet-triplet imbalance by charging the double dot from the joint left lead, $F \approx 1/2$. 

A finite singlet-triplet relaxation reduces the imbalance between singlet and triplet in the stationary state, and the Aharonov-Bohm oscillations are restored. To estimate the influence of spin flip and dephasing we introduce phenomenological transition rates between singlet and triplets.
For simplicity we choose all of them equal to $\Gamma_\mrm{ST}$.
To observe a significant suppression of the Aharonov-Bohm oscillations a relaxation rate smaller than the current rate between source and drain is required, i.e.\@ $\Gamma_\mrm{ST} < \GRud$ in our case.

In Figure \ref{Fig:ABI-Fork_GRcomp_phi} the first and second order conductance contributions are plotted individually. 
The transport supported by the equilibrium state is dominated by the second order triplet contribution and the Aharonov-Bohm oscillations are well visible. The singlet-triplet relaxation does not affect the behavior.
If the electrons are strongly driven into the double dot from the joint left reservoir, a singlet-triplet imbalance forms and the oscillations of the singlet and triplet contributions cancel each other. A finite relaxation rate partially destroys the imbalance and the cancellation. 
The conductance supported by the singly occupied states becomes quite strong for a large bias voltage, however, its phase dependence is weak since for strongly asymmetric coupling the isospin is parallel to $\mbf{m}_\mrm{L}$ with $|\mbf{I}_{\uparrow / \downarrow}| \approx 1/8\,p_1$ such that $g_1 \approx 4\,p_1$.

\section{Conclusions}

Here we proposed a setup for the generation and detection of spin entanglement between two spatially separated electrons. 
A bias voltage applied across a double-dot system with strong onsite Coulomb repulsion, coherently coupled to a joint source electrode and individually coupled to separate drains, creates an imbalance between singlet and triplet probabilities. 
An asymmetry in the coupling of source and drain increases the overall probability to find two excess electrons in the double dot system and for them to form a Werner state.
The underlying mechanism of entanglement generation relies on the specific nonequilibrium situation and is fundamentally different from those schemes in which two electrons are extracted from a singlet ground state.

To detect the singlet-triplet imbalance we added an Aharonov-Bohm probe, obtained by coupling a second joint reservoir coherently to the double dot. The enclosed flux controls the interference pattern which depends on the state of the double dot.
In particular, the cotunneling transport through the Aharonov-Bohm ring is sensitive to an imbalance in the singlet-triplet distribution. 

We compared two situations: 
For vanishing bias voltage between the joint source and the drain reservoirs the double dot is in equilibrium, and singlet and triplet are uniformly distributed, and the Aharonov-Bohm conductance is strongly flux dependent.
On the other hand, by applying a strong bias voltage a singlet-triplet imbalance is generated, and the Aharonov-Bohm oscillations are suppressed. By choosing asymmetric couplings we can ensure that the double dot was predominantly occupied with two electrons.

For an experimental realization one needs a geometry with spatial separation 
of the dots shorter than the phase coherence length.
Finite decoherence due to spin-orbit effects, hyperfine coupling or coupling 
to some other bath will reduce the overall visibility of the Aharonov-Bohm
oscillations.
Nevertheless, the extra suppression of the Aharonov-Bohm oscillations for 
strong bias voltage as compared to equilibrium will indicate a singlet-triplet 
asymmetry.
The probability to find the system occupied with two electrons can be 
enhanced by tuning the dot levels well below the Fermi energy of the probe 
and choosing a strong asymmetry in the coupling of source and drain.
For the detection of a suppressed phase dependence caused by a singlet-triplet 
imbalance a tunneling rate between source and drain much larger 
than the spin decoherence time is required. 

\section*{Acknowledgments}

We acknowledge stimulating discussions with G. Burkard, P.W. Brouwer,
J. Weis, J. Martinek, Y. Gefen, T. L\"ofwander, and
E. Prada. This work was supported by the Landesstiftung
Baden-W\"urttemberg via the Kompetenznetz
Funktionelle Nanostrukturen, and DFG via SFB 491 and SPP 1285.

\end{document}